\documentclass{article}
\usepackage{codefuse_tech_report}
\usepackage[colorlinks = true,
            linkcolor = blue,
            urlcolor  = blue,
            citecolor = blue,
            anchorcolor = blue]{hyperref}   
\usepackage{microtype}
\usepackage{xurl}
\usepackage{booktabs}
\usepackage{enumitem}
\usepackage{multicol}
\usepackage{CJKutf8}
\usepackage{siunitx}
\usepackage{floatflt}
\usepackage{graphicx}
\usepackage{wrapfig}
\usepackage{authblk}
\usepackage{lipsum}
\usepackage{algorithm}
\usepackage{algorithmicx}
\usepackage{algpseudocode}
\usepackage{subfigure}
\usepackage{multirow}
\usepackage{pifont}  

\algnewcommand{\LeftComment}[1]{\Statex \(\triangleright\) #1}

\usepackage{array}
\usepackage{amsmath}
\usepackage{amssymb}
\usepackage{mathtools}
\usepackage{amsthm}

\usepackage[capitalize,noabbrev]{cleveref}
\usepackage{adjustbox} 

\theoremstyle{plain}

\theoremstyle{definition}

\theoremstyle{remark}

\colmfinalcopy

\usepackage[utf8]{inputenc}
\usepackage[T1]{fontenc}
\usepackage{caption} 
\usepackage{adjustbox} 
\usepackage{fontawesome5}

\usepackage{tcolorbox}
\tcbuselibrary{skins,breakable}

\usepackage{tikz}
\usetikzlibrary{math,matrix, calc, shapes, positioning, fit, arrows.meta, backgrounds, positioning}
\usetikzlibrary{arrows}
\usepackage{wrapfig}
\usepackage{lipsum}
\usepackage{amsmath}    
\usepackage{enumitem}
\setlist{leftmargin=5.5mm}

\usepackage{booktabs}
\usepackage{footmisc} 
\usepackage{multirow, multicol}
\usepackage{pifont}       
\usepackage{xcolor}       

\usepackage{pgfplots}
\pgfplotsset{compat=1.18} 

\usepackage{tcolorbox}
\usepackage{xspace}
\usepackage{graphicx}
\usepackage{authblk}
\usepackage[figuresright]{rotating}
\newcommand{\warningsign}{\tikz[baseline=-.75ex] \node[shape=regular polygon, regular polygon sides=3, inner sep=-1pt, draw, thick] {\textbf{!}};}

\usepackage{pifont}
\newcommand{\cmark}{\textcolor{green!80!black}{\ding{51}}}
\newcommand{\xmark}{\textcolor{red}{\ding{55}}}
\newcommand{\warn}{\textcolor{orange}{\warningsign}}
\newcommand{\dmark}{\textcolor{orange}{\ding{108}}}

\newcommand{\rating}[1]{\textcolor{orange!80!black} {\ifcase#1 \or $\star\xspace$ \or $\star\star\xspace$ \or $\star\star\star\xspace$ \or $\star\star\star\star\xspace$ \or $\star\star\star\star\star$ \fi}}

\newcommand{\derisk}{\texttt{OpenDerisk}\xspace}

\tcbuselibrary{skins}

\usepackage{color}
\usepackage{xcolor}
\usepackage{soul} 

\definecolor{tred}{RGB}{251, 130, 132}
\definecolor{torange}{RGB}{247, 162, 116}
\definecolor{tyellow}{RGB}{251, 218, 140}
\definecolor{tgreen}{RGB}{127, 204, 181}
\definecolor{tblue}{RGB}{89, 177, 215}
\definecolor{insightblue}{RGB}{162, 210, 255}
\definecolor{questionred}{RGB}{255, 175, 204}

\title{OpenDerisk: An Industrial Framework for AI-Driven SRE, with Design, Implementation, and Case Studies}

\author{	Peng Di\thanks{Equal Contribution. Correspondence to: Peng Di \textless dipeng.dp@antgroup.com\textgreater ~and Faqiang Chen \textless faqiang.cfq@antgroup.com\textgreater}$^{*}$	}
\author{	Faqiang Chen$^{*}$	}
\author{	Xiao Bai 	}
\author{	Hongjun Yang  	}
\author{	Qingfeng Li 	}
\author{	Ganglin Wei 	}
\author{	Jian Mou 	}
\author{	Feng Shi 	}
\author{	Keting Chen 	}
\author{	Peng Tang 	}
\author{	Zhitao Shen 	}
\author{	Zheng Li 	}
\author{	Wenhui Shi 	}
\author{	Junwei Guo 	}
\author{	Hang Yu	}

\affil{%
    Ant Group, China\\
    \faGithub ~\url{https://github.com/derisk-ai/OpenDerisk}
}

\colmfinalcopy 

\begin{document}

\maketitle

\begin{figure}[h!]
    \centering
    \includegraphics[width=0.9\linewidth]{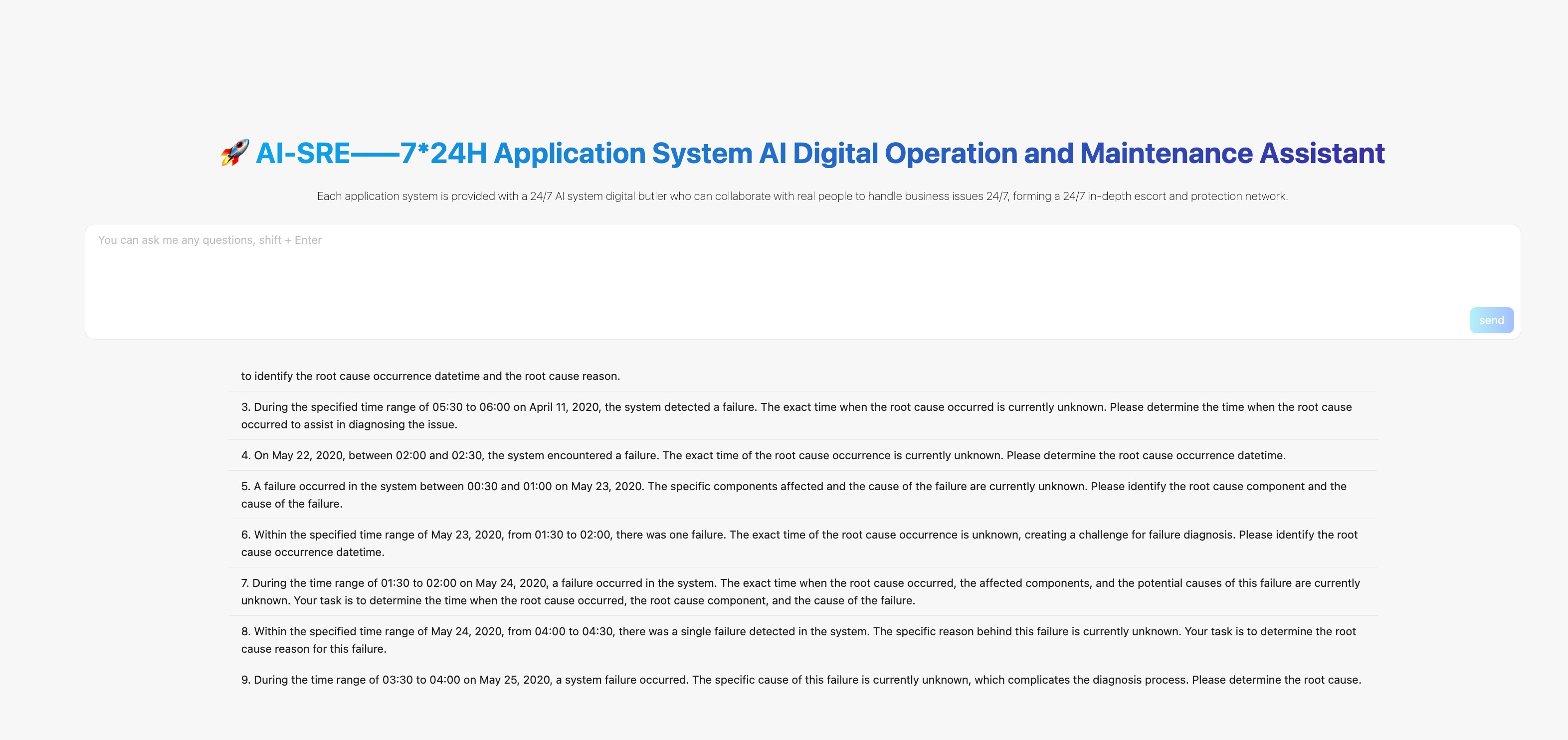}
    \caption{The web-based user interface for \derisk is open source and available at \url{https://github.com/derisk-ai/OpenDerisk/}}
    \label{fig:overview}
\end{figure}


\begin{abstract}
The escalating complexity of modern software imposes an unsustainable operational burden on Site Reliability Engineering (SRE) teams, demanding AI-driven automation that can emulate expert diagnostic reasoning. Existing solutions, from traditional AI methods to general-purpose multi-agent systems, fall short: they either lack deep causal reasoning or are not tailored for the specialized, investigative workflows unique to SRE. To address this gap, we present \derisk, a specialized, open-source multi-agent framework architected for SRE. \derisk integrates a diagnostic-native collaboration model, a pluggable reasoning engine, a knowledge engine, and a standardized protocol (MCP) to enable specialist agents to collectively solve complex, multi-domain problems. Our comprehensive evaluation demonstrates that \derisk significantly outperforms state-of-the-art baselines in both accuracy and efficiency. This effectiveness is validated by its large-scale production deployment at Ant Group, where it serves over 3,000 daily users across diverse scenarios, confirming its industrial-grade scalability and practical impact. \derisk is open source and available at \url{https://github.com/derisk-ai/OpenDerisk/}
\end{abstract}


\section{Introduction}
\label{sec:introduction}

The complexity of modern software---defined by distributed microservices, cloud-native architectures, and relentless release cycles---has surpassed human cognitive scale. For Site Reliability Engineering (SRE) teams, this creates an unsustainable operational burden, making AI-driven automation an operational necessity, not a luxury~\cite{booklu2025}. However, the central challenge is profound: SRE tasks are not simple computations but complex cognitive investigations. Whether performing Root Cause Analysis (RCA) or assessing risk, an engineer must synthesize disparate signals into a coherent causal narrative through hypothesis testing and deep system knowledge. This sophisticated, diagnostic reasoning has eluded previous automation efforts focused on mere pattern matching. This impasse raises a pivotal question: \textbf{can we architect a system that emulates the investigative sense-making of an expert SRE?}

The pursuit of SRE automation has produced generations of tools, each with inherent limitations~\cite{10.5555/2810087}. Traditional learning-based methods, such as causal discovery~\cite{chakraborty2023causil,bi2024faultinsight,arnold2007temporal,li2022causal} and dependency graph analysis~\cite{zheng2024mulan,wang2023incremental}, are often constrained by the immense complexity of real-world systems. More recently, multi-agent frameworks have emerged. The state-of-the-art OpenRCA~\cite{xu2025openrca} uses a multi-agent system for RCA, but its generalist problem-solving model struggles to orchestrate deep, multi-domain expertise and is difficult to extend. Similarly, general-purpose systems, like MetaGPT,  TaskWeaver and others~\cite{hong2024metagpt,zhang2024ufouifocusedagentwindows,qiao2024taskweavercodefirstagentframework}, are fundamentally tailored for \textit{generative} development, not the investigative and diagnostic workflows central to SRE.

To bridge these gaps, we present \textbf{\derisk}, a specialized multi-agent framework designed to augment, not just automate, human expertise. Its architecture is built on four integrated core components: a \textbf{Multi-Agent System} that enables adaptive collaboration patterns; a pluggable \textbf{Reasoning Engine} that endows each agent with multi-step cognitive abilities; a sophisticated \textbf{Knowledge Engine} that grounds their analysis in domain-specific data; and a standardized \textbf{Model Context Protocol (MCP)} that ensures modularity and extensibility.

We validate this architecture through a comprehensive evaluation focused on three key research questions. We first demonstrate that \derisk's multi-agent approach \textbf{significantly outperforms monolithic agents in both accuracy and efficiency (RQ1)}. We then assess its \textbf{adaptability by successfully integrating new knowledge and swapping foundational LLMs (RQ2)}. Finally, an in-depth ablation study quantifies the \textbf{crucial contribution of each architectural component (RQ3)} to the system's overall performance.

This proven effectiveness, combined with its inherent flexibility, allows \derisk to function as a practical and powerful ``co-pilot'' for SRE teams. This is substantiated by its successful production deployment at Ant Group, where, in just three months, it was adopted for \textbf{13 new application scenarios} with over \textbf{50 new specialized agents} created by developers. Today, the platform serves more than \textbf{3,000 daily users} and executes over \textbf{60,000 runs per day}, demonstrating its industrial-grade impact and scalability.

    
    

This paper's primary contributions are as follows:

\begin{itemize}
    \item \textbf{Design:} We present a novel framework architecture for SRE where specialist agents, each equipped with a pluggable reasoning engine and grounded in expert knowledge, collaborate within a diagnostic-native paradigm that is governed by a protocol-driven (MCP) approach.
    
    
    \item \textbf{Implementation:} We detail the implementation of \derisk as a robust, industrial-grade system. This includes key architectural optimizations such as \textit{advanced context engineering} to manage long-running tasks, \textit{sophisticated agent orchestration} for dynamic collaboration, and integrated \textit{visualization and human-in-the-loop features} to enhance user trust and control.

    \item \textbf{Large-Scale Empirical Validation:} We provide extensive evidence of \derisk's effectiveness and scalability. This validation includes quantitative metrics from its successful production deployment at Ant Group, where it serves over \textbf{3,000 daily users} and executes \textbf{60,000+ runs per day}, as well as qualitative insights from in-depth industrial case studies on diverse SRE workflows.

    \item \textbf{Open-Source Contribution:} As a key contribution, we have open-sourced this core framework\footnote{The open-source project is available at: \url{https://github.com/derisk-ai/OpenDerisk}} to foster research and reproducibility.
\end{itemize}
\section{Related Work}

Our work is positioned at the intersection of SRE automation, LLM-based software engineering, multi-agent systems, and context engineering. This section reviews the state of the art in each domain and situates the contributions of \derisk.

\paragraph{SRE Automation.}
The pursuit of automated SRE, often under the umbrella of AIOps, has a rich history focused on mitigating the operational burden of complex systems. Early efforts centered on statistical methods for anomaly detection~\cite{chandola2009anomaly} and log analysis~\cite{he2016experience,aniche-faults-in-webapis,aniche-log-placement-recommendation} to identify deviations from normal behavior. More advanced approaches utilize graph-based structures to model system dependencies and pinpoint root causes. Frameworks like CloudRCA~\cite{li2022cloudrca} and MicroRCA~\cite{wang2023microrca} leverage service dependency graphs and call chains to trace the propagation of failures. RCAEval~\cite{pham2025rcaeval,pham2024baro,pham2024root} offers an open-source benchmark with an evaluation framework for root cause analysis (RCA) in microservice systems. Concurrently, methods based on causal discovery~\cite{spirtes2000causation} attempt to infer the underlying causal graph from observational data to distinguish correlation from causation~\cite{cheng2022causalrca}. 
While powerful at correlating symptoms across vast datasets, these AI methods fundamentally lack deep, semantic reasoning. They can identify anomalous patterns and likely failure paths but struggle to interpret the why behind an incident, a cognitive leap that requires understanding system logic, documentation, and operational context. \derisk bridges this gap by integrating the statistical prowess of these methods with the semantic and causal reasoning capabilities of Large Language Models (LLMs).

\paragraph{LLM-based Software Engineering.}
The advent of LLMs has revolutionized the software development lifecycle (SDLC). Foundational models like Codex~\cite{chen2021evaluating} demonstrated remarkable capabilities in code generation, leading to a new generation of AI-powered software engineering agents. Systems like SWE-agent~\cite{yang2024sweagent} and OpenDevin~\cite{wang2024opendevin} act as autonomous developers, capable of understanding bug reports, navigating codebases, writing code, and executing tests to resolve complex issues. Other works have focused on specialized tasks like automated program repair~\cite{xia2023automated,monperrus2018automatic,zhao2023rightpromptsjobrepair,fan2023static,rondon2025evaluatingagentbasedprogramrepair,zhang2024autocoderover,liu2024marscode} and test-free fault localization~\cite{ni2023leveraging,qin2024agentfl}. This body of work provides compelling evidence that LLMs can perform complex, multi-step tasks within the software domain. However, their focus has overwhelmingly been on generative tasks—creating or modifying code. SRE work, in contrast, is primarily investigative and diagnostic. It requires a different cognitive workflow centered on hypothesis generation, evidence gathering, and logical deduction. \derisk adapts the proven power of LLMs from generative SE to the specific diagnostic challenges inherent in SRE.

\paragraph{Multi-Agent Frameworks.}
As tasks grow in complexity, the "divide and conquer" strategy of multi-agent systems has proven superior to monolithic agent designs~\cite{li2024survey, li2023camel,rizk2020unified,finin1994kqml,dong2024agentopsenablingobservabilityllm}. General-purpose frameworks like AutoGen~\cite{wu2023autogen}, MetaGPT~\cite{hong2024metagpt}, TaskWeaver~\cite{qiao2024taskweavercodefirstagentframework}, SEEAgent \cite{bui2025llmbasedmultiagentframeworkagile} and others orchestrate teams of LLM agents to collaboratively solve problems, from writing entire software projects to complex data analysis. These systems have pioneered sophisticated collaboration patterns, role assignments, and communication protocols. However, their collaboration models are typically optimized for software construction or general problem-solving, not for the structured, high-stakes investigation of a system outage. More domain-specific systems like OpenRCA~\cite{xu2025openrca} apply the multi-agent paradigm to SRE but often employ a generalist problem-solving model that struggles to coordinate the deep, specialized expertise required for industrial incidents. \derisk distinguishes itself by implementing a diagnostic-native collaboration model, where specialized agents (e.g., a "database expert," a "network analyst") work together in a manner that mirrors an expert SRE team during an incident investigation.

\paragraph{Context Engineering and Optimization.}
The performance of any LLM-based system is critically dependent on its ability to manage the finite context window~\cite{pinto2024lessonsbuildingstackspotai,manus}. A significant body of research addresses this challenge. Retrieval-Augmented Generation (RAG)~\cite{lewis2020retrieval} has become the standard for injecting external, up-to-date knowledge into the context. Reasoning techniques like Chain-of-Thought (CoT)~\cite{wei2022chain} and ReAct~\cite{yao2022react} structure the context to elicit more robust, step-by-step reasoning. Furthermore, frameworks like Toolformer~\cite{schick2023toolformer} and Gorilla~\cite{patil2023gorilla} have established methods for teaching LLMs to use external tools via API calls represented in their context. While these are powerful, foundational techniques, they are often applied generically. SRE diagnostics require managing a long-running, evolving context filled with heterogeneous data (logs, metrics, traces, alerts, documentation) and a dynamic set of tools. A simple ReAct loop or RAG query is insufficient. \derisk addresses this with its Model Context Protocol (MCP), a specialized form of context engineering that provides a standardized, stateful structure for managing the complex interplay of observations, thoughts, actions, and tool outputs throughout a protracted diagnostic session.

In summary, \derisk synthesizes advances from these four domains to create a novel solution. It leverages the diagnostic power of LLMs, organizes them within a specialized multi-agent framework tailored for SRE workflows, and governs their operation through a sophisticated context engineering protocol, thereby overcoming the limitations of prior work.

\section{\derisk System}
\label{sec:system}

\begin{figure*}[ht]
    \centering
    \includegraphics[width=0.9\linewidth]{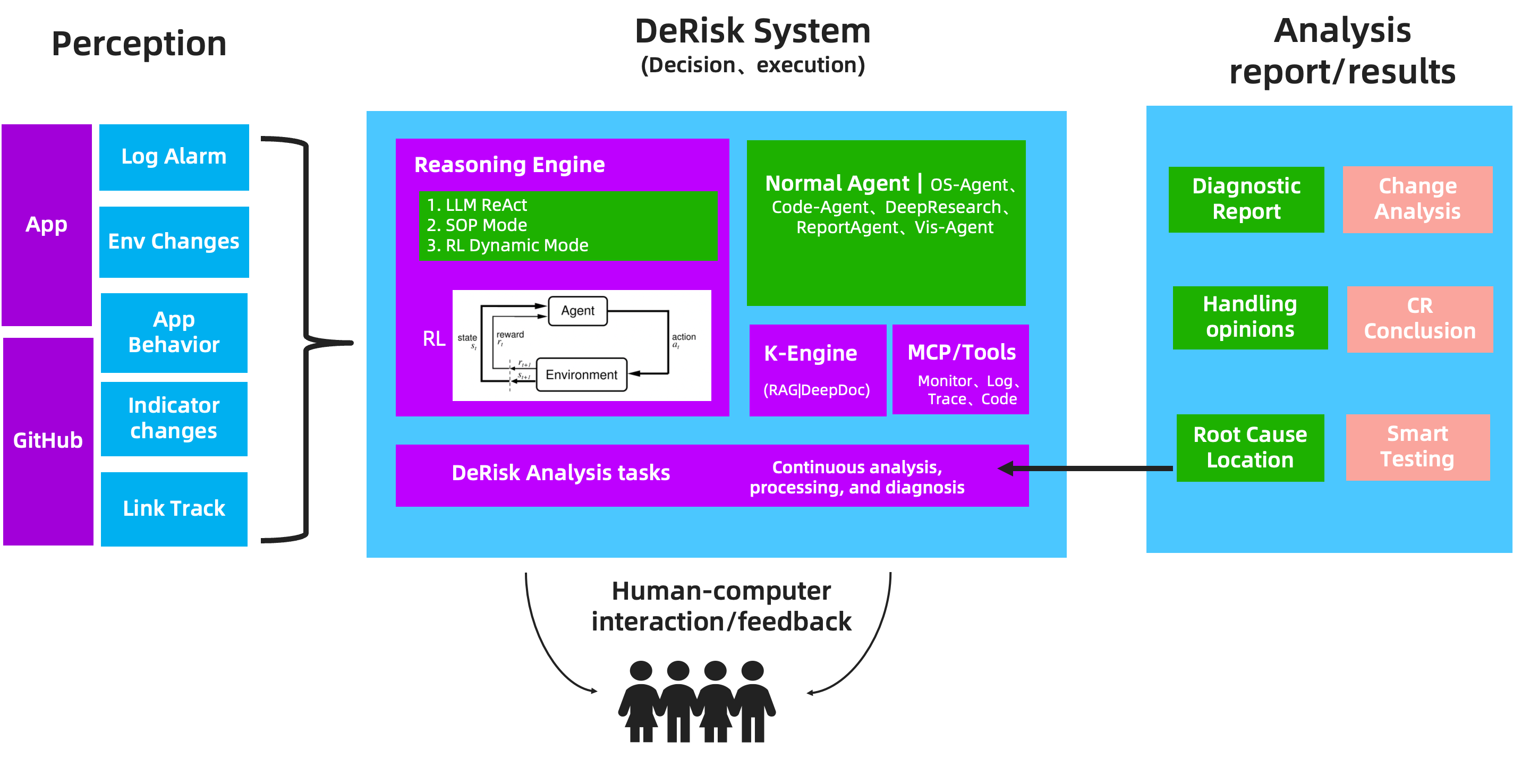} 
    \caption{The end-to-end architecture of the \derisk system, illustrating the flow from Perception to Analysis, centered around the core Decision and Execution engine, and augmented by Human-in-the-Loop feedback.}
    \label{fig:system_architecture}
\end{figure*}

This section details the architecture and core components of the \derisk framework, which is designed as an end-to-end diagnostic pipeline that mimics a human expert's problem-solving process. We first present the design philosophy that underpins the entire system. Then, following the logical flow depicted in Figure~\ref{fig:system_architecture}, we describe the system's three primary stages: a \textit{Perception Layer} to sense the environment, a core \textit{DeRisk System} for decision-making and execution, and an \textit{Analysis Reporting Layer} to deliver actionable insights.

\subsection{Design Philosophy}
The architecture of \derisk is a deliberate choice, resulting from an evaluation of three dominant technical paradigms. We dismissed the traditional \textit{Workflow} paradigm due to its inherent rigidity. While we recognize the immense future potential of \textit{Agentic Reinforcement Learning (Agentic RL)}, its current technical immaturity and high resource requirements make it impractical for present-day deployment. Therefore, \derisk is strategically built upon the \textbf{Multi-Agent ReAct} paradigm. This approach leverages the advanced reasoning of LLMs to create a flexible, intelligent, and collaborative system best aligned with modern engineering principles for building evolvable AI systems.

This philosophy is realized through three primary optimization objectives:
\begin{itemize}
    \item \textbf{Multi-Agent Collaboration Optimization:} To continuously discover and refine the most effective collaboration paradigms among agents.
    \item \textbf{Context Engineering Optimization:} To treat context generation as a formal optimization problem, maximizing the output quality of the LLM.
    \item \textbf{System-Level Reinforcement Learning:} To view the former two as local optimizations, with the ultimate goal of enabling end-to-end systemic evolution through online learning.
\end{itemize}

\subsection{System Workflow and Components}

\subsubsection{Perception Layer: Sensing the Environment}
The system's diagnostic workflow begins at the Perception Layer, which is responsible for ingesting a diverse array of signals that indicate a potential or ongoing incident. These signals are sourced from various operational and development platforms, including: \texttt{Log Alarms}, anomalous \texttt{App Behavior}, \texttt{Environment Changes}, and code changes from \texttt{GitHub}. These perceptual inputs serve as the initial trigger for the core DeRisk System.

\subsubsection{The DeRisk System: Core Decision and Execution}
\begin{figure*}[t]
    \centering
    \includegraphics[width=0.9\linewidth]{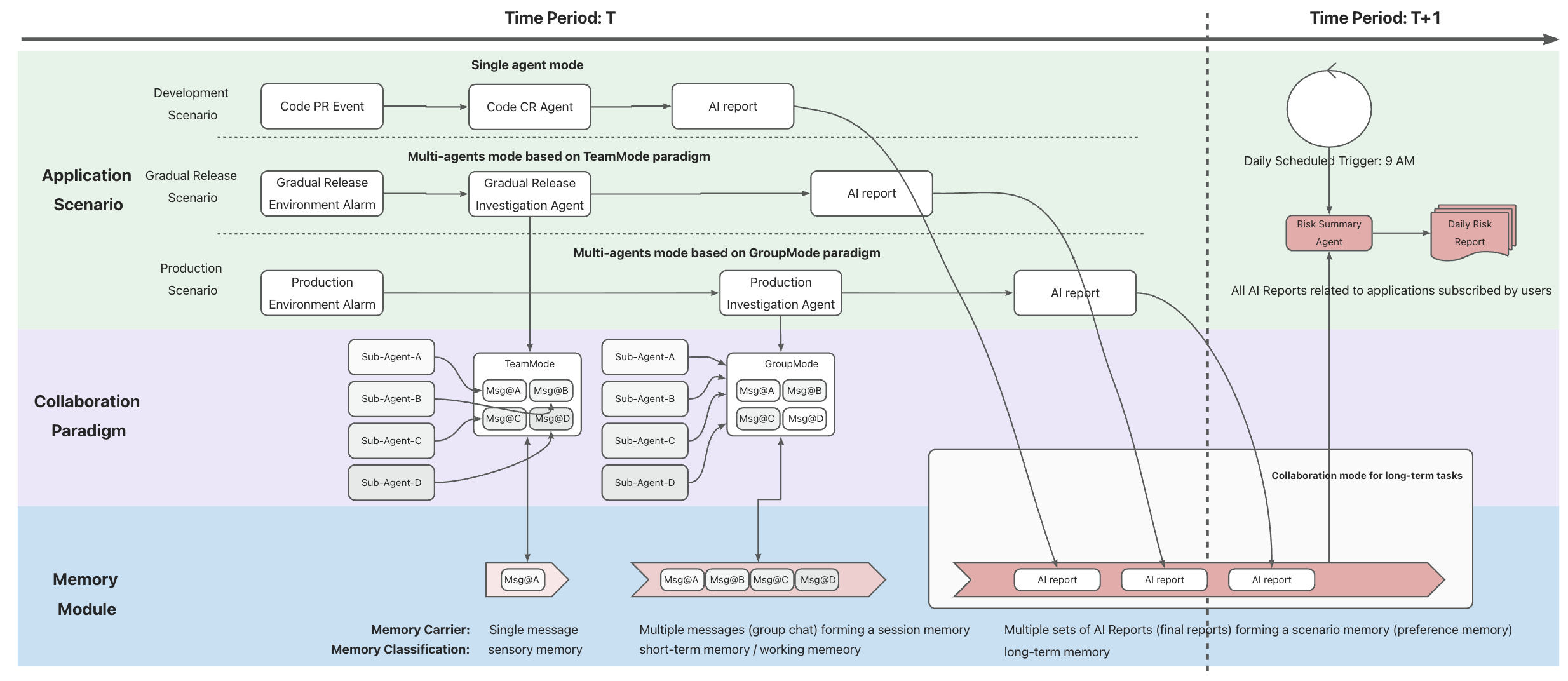}
    \caption{Multi-Agent Collaboration workflow in \derisk. The framework dynamically selects a collaboration paradigm (Single-Agent, TeamMode, or GroupMode) based on the application scenario, supported by a multi-layered memory module.}
    \label{fig:multi_agent_collaboration}
\end{figure*}

The DeRisk System is the central nervous system of the framework, where raw perceptual data is transformed into intelligent decisions and actions. This is where the core \texttt{DeRisk Analysis tasks}—continuous analysis, processing, and diagnosis—take place. It is composed of several tightly integrated components:

\paragraph{Multi-Agent System and Collaboration Paradigms}
\derisk employs a multi-agent framework to enable a sophisticated division of labor. As detailed in Figure~\ref{fig:multi_agent_collaboration}, the framework orchestrates a team of specialized agents (e.g., \texttt{OS-Agent}, \texttt{Code-Agent}) and dynamically adapts its collaboration strategy (e.g., `TeamMode`, `GroupMode`) based on the complexity of the application scenario.

\paragraph{Reasoning Engine}
At the heart of each agent lies a pluggable Reasoning Engine, the cognitive core responsible for planning and decision-making. It supports multiple modes of operation, including dynamic \textbf{LLM ReAct Mode} for exploratory analysis, deterministic \textbf{SOP Mode} (Standard Operating Procedure Mode) for routine tasks, and an \textbf{RL Dynamic Mode} for self-optimization.

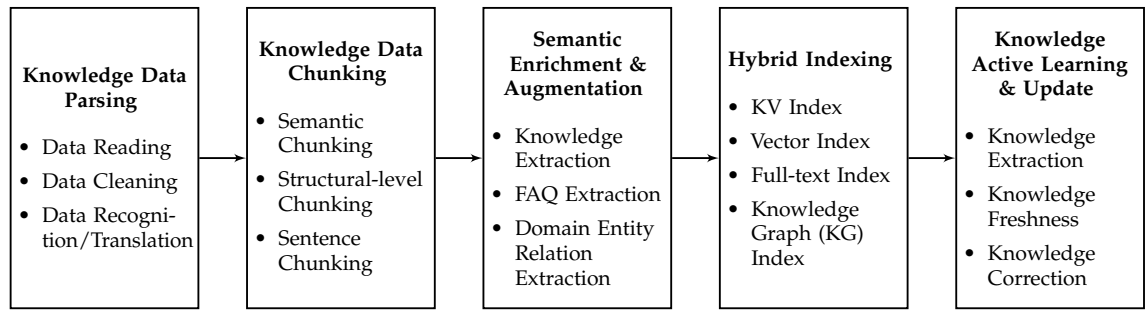
\begin{figure}
    \centering
\resizebox{\columnwidth}{!}{
    \footnotesize
\begin{tikzpicture}[node distance=0.7cm, auto, >=latex']
    \tikzset{
        process_block/.style={
            rectangle, draw, black, 
            fill=white, 
            text=black, 
            thick, 
            minimum width=2.7cm, minimum height=4.5cm, 
            text width=2.6cm, 
            align=left, 
        },
    }
    \node (p1) [process_block] {
        \centering \textbf{Knowledge Data Parsing}
        \vspace{0.2cm} 
        \begin{itemize}[leftmargin=*]
            \item Data Reading
            \item Data Cleaning
            \item Data Recognition/Translation
        \end{itemize}
    };

    \node (p2) [process_block, right=of p1] {
        \centering \textbf{Knowledge Data Chunking}
        \vspace{0.2cm}
        \begin{itemize}[leftmargin=*]
            \item Semantic Chunking
            \item Structural-level Chunking
            \item Sentence Chunking
        \end{itemize}
    };

    \node (p3) [process_block, right=of p2] {
        \centering \textbf{Semantic Enrichment \& Augmentation}
        \vspace{0.2cm}
        \begin{itemize}[leftmargin=*]
            \item Knowledge Extraction
            \item FAQ Extraction
            \item Domain Entity Relation Extraction
        \end{itemize}
    };

    \node (p4) [process_block, right=of p3] {
        \centering \textbf{Hybrid Indexing}
        \vspace{0.2cm}
        \begin{itemize}[leftmargin=*]
            \item KV Index
            \item Vector Index
            \item Full-text Index
            \item Knowledge Graph (KG) Index
        \end{itemize}
    };

    \node (p5) [process_block, right=of p4] {
        \centering \textbf{Knowledge Active Learning \& Update}
        \vspace{0.2cm}
        \begin{itemize}[leftmargin=*]
            \item Knowledge Extraction
            \item Knowledge Freshness
            \item Knowledge Correction
        \end{itemize}
    };

    \draw[->, thick] (p1) -- (p2);
    \draw[->, thick] (p2) -- (p3);
    \draw[->, thick] (p3) -- (p4);
    \draw[->, thick] (p4) -- (p5);

\end{tikzpicture}
}
\caption{The five-stage knowledge processing pipeline of the K-Engine, which transforms raw data into a structured, indexed, and actively maintained knowledge base.}
    \label{fig:knowledge_processing}
\end{figure}

\paragraph{Knowledge Engine (K-Engine)}
To empower agents with deep domain expertise, the framework integrates a powerful \textbf{Knowledge Engine (K-Engine)}. This component implements an advanced Retrieval-Augmented Generation (RAG) system by transforming raw enterprise data into an actionable knowledge base through the five-stage pipeline shown in Figure~\ref{fig:knowledge_processing}.

The pipeline operates as follows:
\begin{enumerate}
    \item \textbf{Knowledge Data Parsing:} The process begins by ingesting data from various sources. This stage involves reading the raw data, performing data cleaning to remove noise and inconsistencies, and using data recognition/translation to standardize diverse formats into a unified representation.
    \item \textbf{Knowledge Data Chunking:} To prepare the data for LLM consumption, the cleaned text is broken down into smaller, semantically coherent units. The engine employs multiple strategies, including \textit{semantic chunking}, \textit{structural-level chunking} (e.g., respecting code blocks or table boundaries), and basic \textit{sentence chunking}.
    \item \textbf{Semantic Enrichment \& Augmentation:} Each chunk is then enriched to improve retrieval relevance. This involves extracting key knowledge, generating potential FAQ pairs, and identifying and tagging domain-specific entities and their relationships.
    \item \textbf{Hybrid Indexing:} Recognizing that no single index is optimal, the engine constructs a hybrid index to support multi-faceted queries. This includes a \textbf{KV Index} for fast lookups, a \textbf{Vector Index} for semantic search, a \textbf{Full-text Index} for precise keyword matching, and a \textbf{Knowledge Graph (KG) Index} to model complex, multi-hop relationships.
    \item \textbf{Knowledge Active Learning \& Update:} Finally, to ensure the knowledge base does not become stale, an active learning loop continuously maintains it. This involves ongoing knowledge extraction from new data, checks for knowledge freshness, and knowledge correction based on feedback or newly discovered information.
\end{enumerate}
This comprehensive process ensures that the K-Engine provides agents with a deep, reliable, and up-to-date understanding of the operational environment.

\paragraph{Tools and Model Context Protocol (MCP)}
Agents interact with the live environment through a standardized set of \textbf{Tools} for tasks such as monitoring and log analysis. The \textbf{Model Context Protocol (MCP)} is a formal contract that defines how tools are described and used, ensuring the system is highly extensible.

\subsubsection{Analysis and Reporting Layer}
The final stage of the pipeline is the Analysis and Reporting Layer, which synthesizes the system's findings into human-readable outputs. This layer produces artifacts such as \texttt{Diagnostic Reports}, \texttt{Root Cause Location}, and \texttt{Handling opinions}.

\subsubsection{Human-in-the-Loop Feedback}
A cornerstone of the \derisk design is the continuous Human-in-the-Loop (HITL) feedback mechanism. At any point, an SRE can interact with the system to provide guidance or corrections. This ensures safety and provides valuable data for the RL training loop, allowing the system to learn from expert intervention.

\section{Implementation}
\label{sec:implementation}

We have implemented the \derisk framework as a robust and scalable system using a standard Python-based ecosystem. The entire framework is open-sourced to encourage further research and community contribution. Our implementation primarily utilized the \texttt{Deepseek}, \texttt{Qwen}, \texttt{Claude} and \texttt{GPT} models. This section details the key engineering decisions and components of our implementation, focusing on our context engineering strategies, agent orchestration, and visualization front-end.

\subsection{Context Engineering Engine}
The finite context window of LLMs is a primary bottleneck for complex, multi-step agent reasoning~\cite{pinto2024lessonsbuildingstackspotai,manus,mei2025surveycontextengineeringlarge}. Our implementation addresses this through a sophisticated context engineering engine designed to maintain a high-signal, low-noise context. This engine is built on three core mechanisms:

\begin{itemize}
    \item \textbf{Summary-Based Memory Compression:} As an agent's working memory (i.e., the history of its actions and observations) grows, the engine employs a compression strategy. Older turns in the conversation history are distilled into concise summaries, preserving key insights while freeing up token space. This allows the agent to maintain a long-term understanding of the task without exceeding the context limit.

    \item \textbf{Configurable Context Policies:} Recognizing that not all context is equally important, the engine supports configurable policies for context construction. An administrator can define rules that dictate how different parts of the context are handled. For example, a policy might specify that a user's original query and the agent's final reasoning steps are preserved in full, while intermediate raw tool outputs are heavily truncated or summarized. This provides granular control over the context payload.

    \item \textbf{Structured Report Distillation:} To prevent information distortion as reports from sub-agents are passed up to a supervisor (a common issue in multi-agent systems), our framework implements a structured distillation process. Instead of passing a full, verbose output, a sub-agent generates a structured summary object. This object contains key findings, confidence scores, and pointers to critical evidence, ensuring the supervisor receives a high-signal, low-noise summary and mitigating the risk of ``lost-in-the-middle'' errors.
\end{itemize}
Together, these mechanisms transform the context from a static transcript into a dynamically managed workspace, enabling sustained reasoning and the ability to perform cognitive refactoring by revising strategy based on a condensed, relevant history.

\subsection{Agent Orchestration}
The framework's logic layer is implemented as a multi-agent architecture orchestrated by a central \texttt{Orchestrator} module. This module manages agent lifecycles, message passing, and task delegation. Each specialized agent (e.g., \texttt{SRE-Agent}, \texttt{Code-Agent}, \texttt{Data-Agent}) is implemented as a class inheriting from a base \texttt{Agent} class, which standardizes its core components: a \textbf{Profile} defining its role and expertise, a set of \textbf{Tools} representing its capabilities, a \textbf{Planning} engine (LLM-powered) for reasoning, a \textbf{Memory} module for maintaining state and context. Inter-agent communication is handled via an asynchronous message bus, allowing for parallel and scalable task execution. The entire system is designed for extensibility, allowing new agents to be seamlessly integrated to tackle emerging SRE challenges.

\begin{table}[]
    \centering
    \scriptsize
    \begin{tabular}{ccc}
\toprule
\textbf{Agent Type} & \textbf{Primary Responsibility} & \textbf{Key Capabilities} \\
\midrule
\textbf{SRE-Agent} & Site Reliability Engineering and orchestration & Incident coordination, system monitoring, agent orchestration \\
\midrule
\textbf{Code-Agent} & Dynamic code generation and analysis & Runtime code generation, static analysis, code-based diagnostics \\
\midrule
\textbf{Data-Agent} & Data processing and analysis & Log analysis, trace processing, metrics evaluation \\
\midrule
\textbf{Vis-Agent} & Visualization and evidence presentation & Evidence chain visualization, diagnostic flow rendering \\
\midrule
\textbf{ReportAgent} & Report generation and documentation & Diagnostic report creation, findings summarization \\
\bottomrule
\end{tabular}

    \caption{
     Roles and capabilities of default digital employees (agents) in \derisk
    }
    \label{tab:agents}
\end{table}

\begin{description}
    \item[Case Study: How to Build an New Agent] 
\end{description}

We will use a real-world industrial case study "Emergency Time-Series Trend Analysis" to illustrate the core concepts, from initial design to a sophisticated, multi-agent validation system. 



\noindent\textbf{The Business Problem:} During incident response, SREs heavily rely on monitoring time-series curves to determine if a system's state is improving, worsening, or stable. The goal was to create a \derisk's agent that can analyze a given monitoring curve and autonomously determine.

\begin{figure}
    \centering
    \includegraphics[width=1\linewidth]{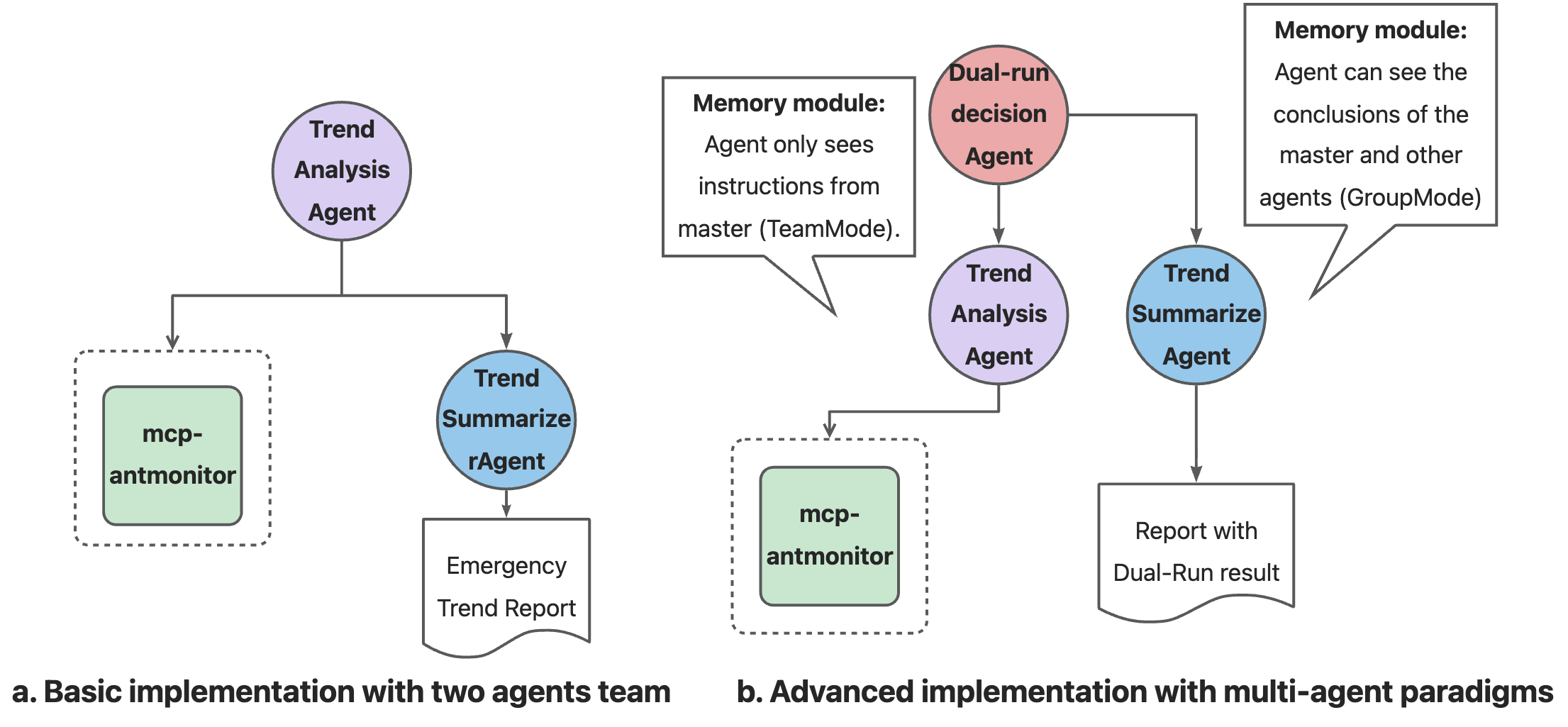}
    \caption{The designs of Emergency Time-Series Trend Agent. Left is basic version with a two-agents team, right is an advanced design with multi-agents paradigm.}
    \label{fig:placeholder}
\end{figure}



The initial implementation focuses on creating a two-agent team: an \texttt{Analysis\_Agent} to perform the core task and a \texttt{Summarizer\_Agent} to format the final report.

\paragraph{Step 1: Building the \texttt{Summarizer\_Agent}}
We start with the simplest component. This agent's sole purpose is to take the findings from other agents and generate a well-formatted report.
\begin{itemize}
    \item \textbf{Agent Type:} `Task-based Agent`.
    \item \textbf{Reasoning Engine:} We select the \texttt{Summarizer Reasoning Engine} in the \derisk tools lib, which is optimized for digesting context and generating reports.
    \item \textbf{Tools \& Knowledge:} None required.
    \item \textbf{System Prompt:} A prompt is configured to define the desired output format, instructing the agent to create a clear, concise report based on the upstream analysis.
\end{itemize}

\paragraph{Step 2: Building the \texttt{Analysis\_Agent} (The ``Master'')}
This agent is responsible for the main logic.
\begin{itemize}
    \item \textbf{Agent Type:} `Task-based Agent` with the \texttt{Default Reasoning Engine}.
    \item \textbf{Tools:} We bind the \texttt{mcp-antmonitor} tool, giving the agent the ability to fetch time-series data from the monitoring system.
    \item \textbf{Sub-Agents:} We link the \texttt{Summarizer\_Agent} created in the previous step.
    \item \textbf{System Prompt (Workflow Definition):} The prompt instructs the agent on its workflow: (1) Receive an alert and time window; (2) Use the \texttt{mcp-antmonitor} tool to retrieve data; (3) Analyze the trend; (4) Pass the data and conclusion to the \texttt{Summarizer\_Agent}.
\end{itemize}

\paragraph{Step 3: Testing and Deployment (V1)}
We test the system with a real-world task query:

\noindent\begin{tcolorbox}[size=title, opacityfill=0.1, nobeforeafter]
``Alert on \texttt{anonymousapp} for 'error rate'. Start time: 2025-08-19 15:21:00. Analyze the monitoring curve trend as of 15:26:00. Is it worsening or recovering?''
\end{tcolorbox}

\begin{figure}
    \centering
    \includegraphics[width=1\linewidth]{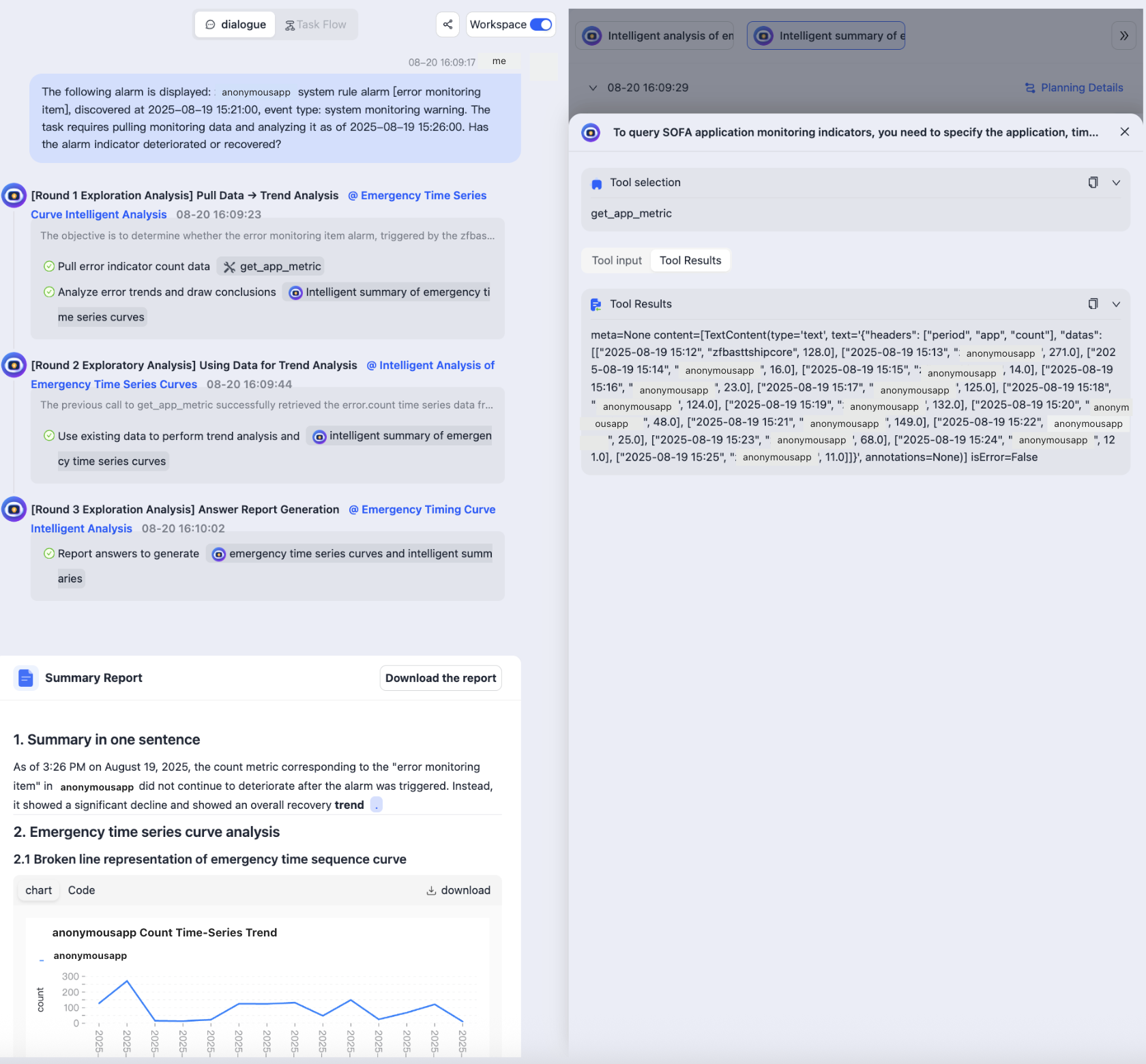}
    \caption{Test Agent—left panel: query and workflow; right panel: sequential execution results. Displayed is the return value from \texttt{mcp-antmonitor}'s \texttt{get\_app\_metric}.}
    \label{fig:testagent}
\end{figure}

The agent successfully fetches the data, correctly identifies the trend, and generates a report with a plotted curve, fulfilling the initial goal, as shown in Figure~\ref{fig:testagent}. That means the agent works. 

\paragraph{Step 4: Building an Advanced Agent}
To validate our agent, we designed an advanced dual-run agent to perform a non-biased comparison against a legacy system's conclusions proposed by basic version. The core challenge is to prevent our new \texttt{Analysis\_Agent} from seeing the legacy answer, which would bias its analysis. This is achieved by leveraging \derisk's advanced features, including a three-agent architecture orchestrated by a \texttt{DualRun\_Orchestrator}. This orchestrator provides a sanitized task to the \texttt{Analysis\_Agent}, which is placed in an isolated \textbf{\texttt{TeamMode}} memory to perform a blind analysis. Finally, a \texttt{Critic\_Summarizer\_Agent} operating in \textbf{\texttt{GroupMode}} gains access to the full context---including both the legacy and new conclusions---to generate a fair comparison report, thus enabling a robust, industrial-grade validation workflow.

\subsection{Visualization and Human-in-the-Loop Interaction}
\begin{figure}[t!]
    \centering
    \includegraphics[width=1\linewidth]{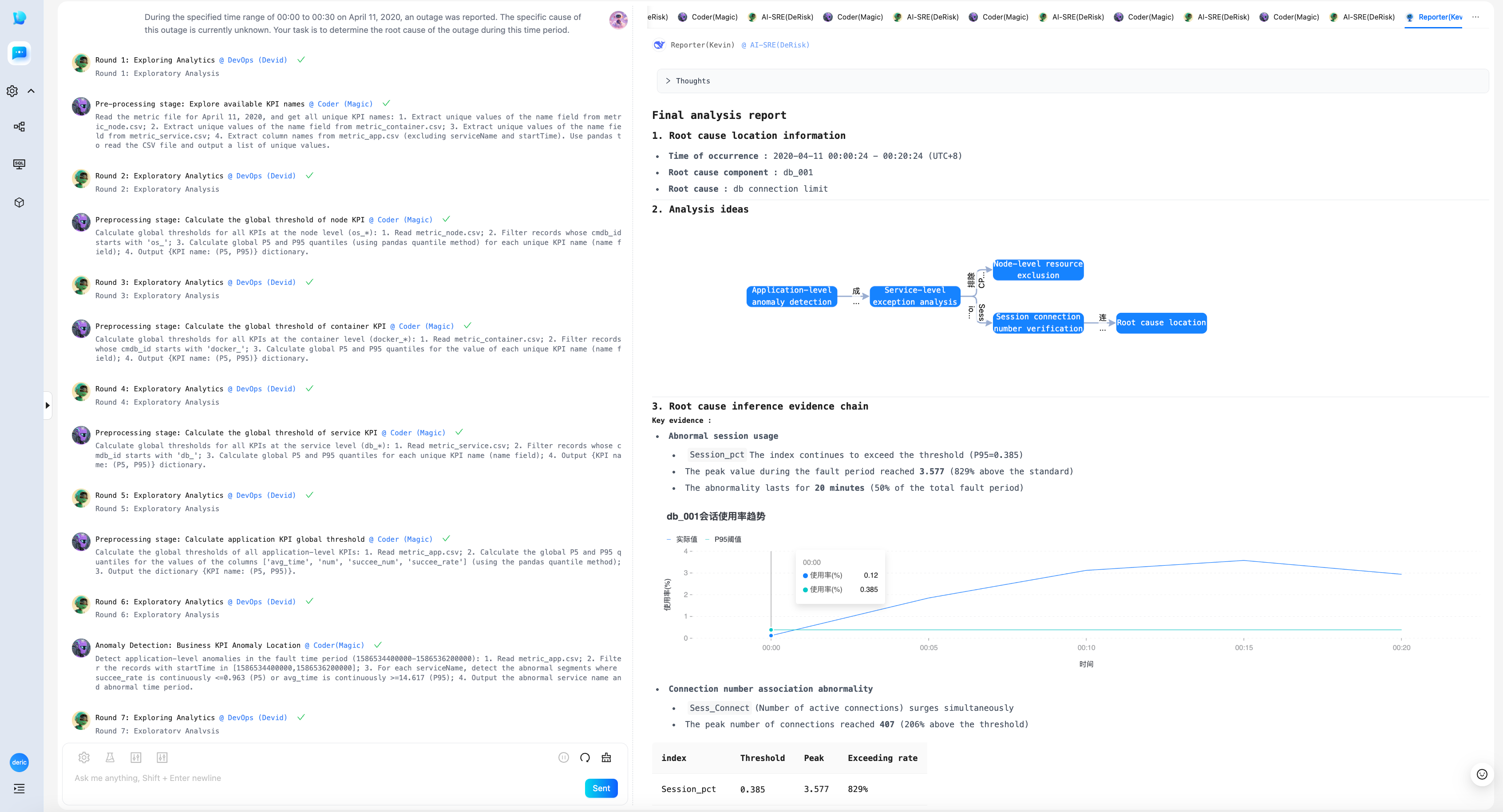}
    \caption{The user interface of \derisk, visualizing the agent's multi-step reasoning and collaboration process.}
    \label{fig:talk}
\end{figure}

To provide critical transparency into the framework's complex internal processes, \derisk includes a web-based user interface that serves as a real-time visualization front-end (Figure~\ref{fig:talk}). This interface is not merely a static dashboard; it is powered by a custom \emph{Visualization Protocol} designed to stream and render the entire problem-solving journey dynamically.

Leveraging this protocol, the UI makes the agent's ``thinking'' process tangible by rendering a comprehensive, step-by-step view of:
\begin{itemize}
    \item \textbf{The Reasoning Flow:} The logical path the supervisor agent takes to decompose and solve the problem.
    \item \textbf{The Evidence Chain:} The specific data (alerts, logs, traces) collected and analyzed by each specialized agent.
    \item \textbf{Multi-Agent Dynamics:} The explicit collaboration and role-switching between different agents as the investigation progresses.
\end{itemize}
This dynamic visualization is essential for building user trust and enabling human-in-the-loop oversight. It allows SREs to not only see the final conclusion but to understand how the system arrived at it, making the AI's diagnostic process fully auditable and intelligible.

\subsection{Deployment at Ant Group}
\label{subsec:deployment}

To validate its practicality and scalability in a real-world setting, the \derisk framework has been deployed in production at Ant Group. The deployment validates a core design tenet of the framework: to serve as a general-purpose AI-SRE platform that empowers developers to autonomously create new agents for diverse SRE tasks, extending far beyond the initial focus on Root Cause Analysis.

The results from this deployment underscore the framework's real-world impact and rapid adoption. Over a three-month production period:
\begin{itemize}
    \item \textbf{Extensibility and Adoption:} The framework has been applied to \textbf{13 new application scenarios}. Internal developers have successfully created over \textbf{50 new specialized agents} to address their unique operational needs, confirming the framework's ease of extension.
    \item \textbf{Scale and Usage:} The platform has achieved significant scale, serving over \textbf{3,000 daily active users} and executing more than \textbf{60,000 diagnostic runs per day}.
\end{itemize}

This large-scale, sustained usage in a mission-critical environment demonstrates that \derisk is not merely a research prototype but a robust, scalable, and valuable industrial-grade system. Building on the success of this internal system, the core technology has been productized into a commercial-grade platform.

\section{Evaluation}
\label{sec:evaluation}

\subsection{Methodology}

Our evaluation is structured around three central research questions (RQs) designed to assess the framework's key attributes, from its overall efficiency to the contribution of its core architectural components. To answer these questions, we compare three primary evolutionary stages of our agent framework, each designed to overcome the limitations of its predecessor.

The Research Questions are as follows:
\begin{itemize}
    \item \textbf{RQ1 (Efficiency and Effectiveness):} \textit{How does our multi-agent architecture compare to monolithic agents in terms of accuracy and execution time on complex industrial tasks?}
    \item \textbf{RQ2 (Adaptability):} \textit{How effectively does the framework integrate new, domain-specific knowledge and adapt to different foundational LLMs?}
    \item \textbf{RQ3 (Ablation and Contribution):} \textit{What is the specific contribution of each architectural enhancement to the agent's performance?}
\end{itemize}

To address these questions, we evaluate the following system configurations:

\paragraph{V1: Basic ReAct Agent.} This baseline represents a single agent operating in a simple \texttt{Think -> Act -> Observe} loop. While capable of basic tool use, this architecture lacks robust end-to-end control and is prone to deviating from the main task when faced with complexity.

\paragraph{V2: Phased-Control ReAct Agent.} This is a more structured, yet still monolithic, single-agent architecture. It introduces a predefined sequence of problem-solving stages (e.g., \textit{Root Cause Analysis}, \textit{Internal Cause Analysis}) and uses a Dynamic Prompt Control mechanism to constrain the agent's behavior at each phase. However, this rigid structure can be brittle to long-chain hallucinations and is sensitive to context window limitations.

\paragraph{V3: Multi-Specialist Agent Framework (Ours).} This is our proposed ``divide and conquer'' paradigm. A central \textbf{Supervisor} agent decomposes the primary task. It then delegates sub-tasks to the most appropriate component: it uses \texttt{ToolCall} to trigger a \textbf{Workflow Engine} for deterministic tasks and \texttt{Handoff} to dispatch complex problems to specialized \textbf{Sub-Agents}. This design dramatically reduces the reasoning burden on any single agent and improves accuracy by leveraging targeted expertise. By comparing V3 against V1 and V2, we can quantify the performance gains and isolate the contribution of each architectural evolution.



\subsection{Case Study 1: Root Cause Analysis}
\label{sec:casestudy_rca}


\subsubsection{Experiment Design}
To empirically evaluate our architectural progression, we designed a comparative experiment measuring the trade-offs between task success and execution time. We compare three distinct agent architectures, each powered by a diverse set of Large Language Models (LLMs) to validate model-agnosticism.


    
    

To demonstrate the generality of our architectural improvements, each configuration was tested with a diverse set of foundational LLMs, including \texttt{Qwen-QWQ-32B}~\cite{qwen2.5,qwq32b}, \texttt{Deepseek-R1-0528}~\cite{deepseekai2025deepseekr1incentivizingreasoningcapability}, and \texttt{Bailing-Deepseek-V3}~\cite{Di_2024,codefuse2025samplemattersleveragingmixtureofexperts} \footnote{\texttt{Bailing-Deepseek-V3} is an in-house model developed by Ant Group's CodeFuse team by fine-tuning Deepseek-V3.}. 


We evaluated all methods on a hybrid benchmark combining the public \textbf{OpenRCA} dataset (335 scenarios) and a proprietary \textbf{AntRCA} dataset. The AntRCA dataset was curated from real-world cases at Ant Group and its composition was adjusted to mirror the true distribution of daily SRE tasks, ensuring high industrial relevance. Performance was quantified using a human-graded 100-point scoring rubric. 


\subsubsection{Results and Analysis}

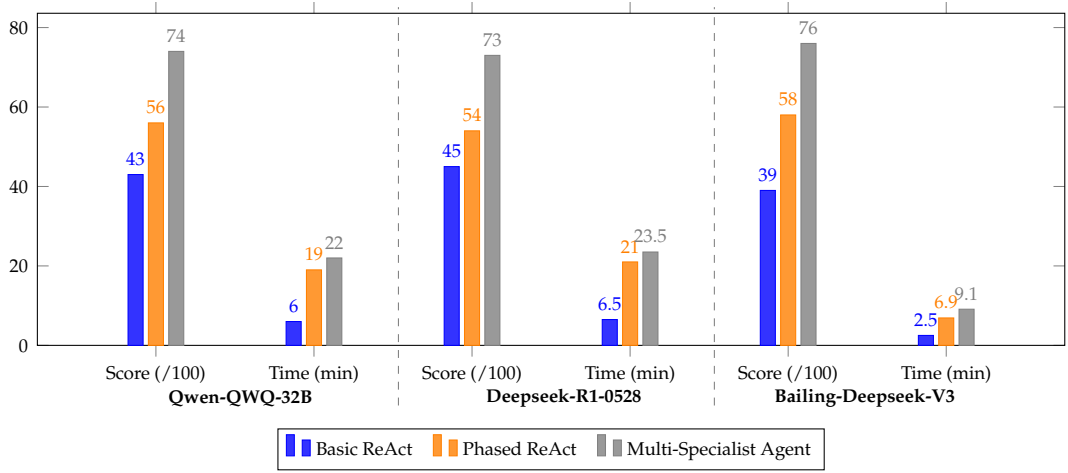
\begin{figure}
\centering
\begin{tikzpicture}
\pgfplotscreateplotcyclelist{mycolors}{
  {blue,fill=blue!80!white},
  {orange,fill=orange!80!white},
  {gray,fill=gray!80!white},
}

\begin{axis}[
    ybar,
    bar width=0.2cm,
    enlarge x limits=0.15, 
    height=6cm,
    width=1\linewidth,
    legend style={
        at={(0.5,-0.25)}, 
        anchor=north,
        legend columns=-1, 
        /tikz/every even column/.append style={column sep=0.2cm}, 
        font=\scriptsize
    },
    ylabel style={font=\bfseries},
    ymin=0, 
    symbolic x coords={
        qwen_score, qwen_time,
        deepseek_score, deepseek_time,
        bailing_score, bailing_time
    },
    xtick=data, 
    xticklabels={
        Score (/100), Time (min),
        Score (/100), Time (min),
        Score (/100), Time (min)
    },
    xticklabel style={font=\scriptsize},
    yticklabel style={font=\scriptsize},
    nodes near coords,
    nodes near coords style={font=\scriptsize},
    cycle list name=mycolors,
]

\addplot coordinates {
    (qwen_score, 43)
    (qwen_time, 6)
    (deepseek_score, 45)
    (deepseek_time, 6.5)
    (bailing_score, 39)
    (bailing_time, 2.5)
};

\addplot coordinates {
    (qwen_score, 56)
    (qwen_time, 19)
    (deepseek_score, 54)
    (deepseek_time, 21)
    (bailing_score, 58)
    (bailing_time, 6.9)
};

\addplot coordinates {
    (qwen_score, 74)
    (qwen_time, 22)
    (deepseek_score, 73)
    (deepseek_time, 23.5)
    (bailing_score, 76)
    (bailing_time, 9.1)
};

\legend{Basic ReAct, Phased ReAct, Multi-Specialist Agent}


\end{axis}

\draw [gray, dashed] (4.8, -0.8) -- (4.8, 4.5);
\draw [gray, dashed] (9, -0.8) -- (9, 4.5);
\node [font=\bfseries\scriptsize]  at (2.7,-0.7) {Qwen-QWQ-32B};
\node [font=\bfseries\scriptsize]  at (7,-0.7) {Deepseek-R1-0528};
\node [font=\bfseries\scriptsize]  at (11,-0.7) {Bailing-Deepseek-V3};


\end{tikzpicture}
\caption{Performance Comparison of Agent Frameworks Across Different Models.}\label{fig:framework_comparison}
\end{figure}


Experimental results, shown in Figure~\ref{fig:framework_comparison},  confirm a consistent performance hierarchy (V3 > V2 > V1) across all tested LLMs, demonstrating the model-agnostic architectural superiority of our Multi-Specialist Agent framework.

\paragraph{Improvement in Task Accuracy}
Across all three base models, \texttt{Qwen-QWQ-32B}, \texttt{Deepseek-R1-0528}, and \texttt{Bailing-Deepseek-V3}, the V3 framework consistently achieved the highest scores. For instance, with the \texttt{Bailing-Deepseek-V3} model, the Multi-Specialist Agent scored \textbf{76}, a substantial improvement over the Phased ReAct's score of \textbf{58} and the Basic ReAct's score of \textbf{39}. This significant leap in performance is attributed to the \textit{divide and conquer} strategy of the V3 architecture. The \texttt{Supervisor} agent effectively decomposes the complex problem into smaller, manageable sub-tasks, which are then handled by specialized agents. This stands in stark contrast to the V1 agent's monolithic and often unfocused reasoning, and the V2 agent's rigid, sequential process, which is vulnerable to cascading errors.

\paragraph{The Accuracy-Efficiency Trade-off}
This improvement in accuracy, however, comes at the cost of increased execution time. The results show that the V3 framework is consistently the slowest, while the V1 framework is the fastest. For example, using \texttt{Qwen-QWQ-32B}, the execution time increased from a mere \textbf{6 minutes} for the V1 agent to \textbf{22 minutes} for the V3 framework. This latency is an expected consequence of the V3 architecture's increased complexity, which involves additional reasoning steps for task decomposition by the \texttt{Supervisor} and inter-agent communication overhead. The V2 architecture represents a middle ground in both accuracy and time.



\paragraph{Conclusion for RQ1}
In conclusion, the Multi-Specialist Agent framework successfully navigates the accuracy-efficiency frontier. It demonstrates that by investing more computational time in a structured, multi-agent reasoning process, it can achieve a dramatic and previously unattainable level of accuracy on complex industrial tasks. The increased latency is a direct and worthwhile trade-off for this substantial gain in problem-solving capability.

\subsection{Case Study 2: Trace Analysis}





\subsubsection{Experimental Design}
This experiment is designed to directly address \textbf{RQ2: How effectively does the framework integrate new, domain-specific knowledge and adapt to different foundational LLMs?} To answer this, we structured our evaluation to systematically test both aspects of the question: knowledge integration and model adaptability.

\begin{itemize}
    \item \textbf{Validating Domain-Specific Knowledge Integration:}
    We first addressed the ``Knowledge Gap''---the absence of Ant Group's proprietary operational knowledge in public LLMs. Using the \derisk framework, we rapidly developed a new trace-based RCA agent by injecting this expert knowledge. To measure the effectiveness of this integration, the agent was deployed into production at Ant Group for a \textbf{one-month trial}. Its performance was validated by \textbf{1,743 developers} across over \textbf{6,000 real-world cases}, providing a robust, in-situ test of its practical capability.

    \item \textbf{Assessing Adaptability to Foundational LLMs:}
    Next, we tested the framework's model-agnosticism. We leveraged the \derisk architecture to systematically integrate and evaluate a diverse set of foundational LLMs (as detailed in Table~\ref{tab:rca_model_comparison}). This comparative analysis on a curated dataset allowed us to quantify how well the framework adapts to different models and identify which are best suited for the task.
\end{itemize}

\paragraph{Evaluation Metrics}
To quantify the outcomes for both objectives, performance was measured across three axes:
\begin{itemize}
    \item \textbf{Aggregate Success Rate:} The percentage of cases where the agent correctly identified the root cause, directly measuring the effectiveness of knowledge integration (Target: >80\%).
    \item \textbf{User-Rated Analysis Quality:} A 1-to-5 star rating from developers, reflecting the perceived utility and quality of the agent's analysis.
    \item \textbf{System Execution Time:} The end-to-end latency, serving as a critical metric for comparing the operational efficiency of different LLMs.
\end{itemize}

\begin{table}[h!]
\centering
\caption{Performance Comparison of LLMs Integrated via the Derisk Framework for Trace-based RCA}
\label{tab:rca_model_comparison}
\begin{tabular}[width=\linewidth]{p{3cm}p{3cm}p{2cm}<{\centering}p{2cm}<{\centering}p{2cm}<{\centering}}
\toprule
\multirow{2}{*}{\textbf{Model}} & \multirow{2}{*}{\textbf{Model Type}} & \textbf{Correct Rate} & \textbf{Analysis} & \multirow{2}{*}{\textbf{Time}} \\ 
& & \textbf{(>80\%)} & \textbf{Quality} & \\ 

\midrule
\texttt{GPT-oss-120B} & Reasoning & \cmark & \rating{5} & Fast \\
\texttt{GPT-oss-20B} & Reasoning & \cmark & \rating{5} & Fast \\
\texttt{GPT-4o} & Non-Reasoning & \cmark & \rating{5} & Very Fast \\
\texttt{Deepseek-v3} & Non-Reasoning & \cmark & \rating{5} & Fast \\
\texttt{Deepseek-r1} & Reasoning & \xmark & \rating{2} & Very Slow \\
\texttt{Kimi-K2} & Non-Reasoning & \cmark * & \rating{4} & Fast *\\
\texttt{Kimi-K1.5} & Non-Reasoning & \cmark & \rating{4} & Slow \\
\texttt{QWQ-32B} & Reasoning & \cmark & \rating{3} & Very Slow \\
\texttt{Qwen3-30B} & Non-Reasoning & \xmark & \rating{1} & Fast \\
\texttt{Qwen3-235B} & Non-Reasoning & \cmark & \rating{5} & Fast \\
\texttt{Qwen3-Coder} & Non-Reasoning & \xmark & \rating{1} & Fast \\
\texttt{Qwen-2.5} & Non-Reasoning & \xmark & \rating{1} & Fast \\
\texttt{Gemini-pro-2.5} & Reasoning & \cmark & \rating{4} & Slow \\
\texttt{Gemini-flash-2.5} & Non-Reasoning & \cmark & \rating{4} & Slow \\
\texttt{Claude-Sonnet-3.5} & Non-Reasoning & \cmark & \rating{5} & Fast \\
\bottomrule
\multicolumn{5}{p{0.9\textwidth}}{\footnotesize * denotes Kimi-K2, which exhibited inconsistent performance across multiple runs. Its correctness rate fluctuated above and below 80\%, with execution times occasionally exceeding 30 seconds.}
\end{tabular}
\end{table}





\subsubsection{Results and Analysis}

The results in Table~\ref{tab:rca_model_comparison} provide the answer to RQ2 by demonstrating the \derisk framework's dual capabilities: effective integration of domain-specific knowledge and seamless adaptability across diverse foundational LLMs.

\paragraph{Effective Knowledge Integration}
The primary success metric---an \textbf{aggregate success rate exceeding 80\%} across 6,000+ cases in the one-month production trial---serves as direct validation for the first part of RQ2. This result proves that the framework successfully bridged the ``Knowledge Gap.'' By providing the tooling to rapidly inject proprietary knowledge from Ant Group, \derisk transformed general-purpose LLMs into high-performing, specialized agents. The high success rate and top-tier \textit{Analysis Quality} scores for models like \textbf{\texttt{GPT-4o}} and \textbf{\texttt{Deepseek-v3}} show that this injected knowledge translates directly into practical, production-grade problem-solving capability.

\paragraph{Model Adaptability and Strategic Tunability}
The breadth of models evaluated in Table~\ref{tab:rca_model_comparison} provides a clear answer to the second part of RQ2. The framework's model-agnostic, ``plug-and-play'' architecture was instrumental in this large-scale comparative analysis. This adaptability is not merely a technical feature but a crucial enabler of strategic tunability. It allows for empirical, data-driven decisions that balance performance, cost, and latency.

For instance, the benchmark data clearly illuminated critical trade-offs: while \texttt{Deepseek-R1} failed to meet performance requirements due to its slow speed, the faster \texttt{Deepseek-v3} emerged as a superior production candidate without sacrificing quality. This data-driven insight, enabled by the framework, allowed us to confidently select \texttt{Deepseek-v3} for the production trial---a decision ultimately validated by the high user-rated quality scores from our 1743 developers. Furthermore, the framework facilitated the identification of model-specific weaknesses, such as the difficulty some \texttt{Qwen-series} models had with JSON parsing, effectively de-risking the model selection process.

\paragraph{Conclusion for RQ2}
In conclusion, this comprehensive evaluation demonstrates that \derisk excels in its core design goals. It serves as a robust meta-platform that provides both the agility to build domain-specific agents and the model-agnostic flexibility to empirically evaluate, select, and deploy them in a demanding industrial environment.

\subsection{Ablation Study}
\label{sec:ablation_study}

\begin{sidewaystable}[thp]
\centering
\scriptsize
\caption{Comprehensive Qualitative Comparison of Agent Architectures (V1-V3)}
\label{tab:comprehensive_comparison}
\begin{tabular}{@{} l l l l l @{}}
\toprule
\textbf{Case Description} & \textbf{Dimension} & \textbf{V1: Basic ReAct} & \textbf{V2: Phased-Control ReAct} & \textbf{V3: Multi-Specialist} \\
\midrule

\multirow{5}{*}{\parbox{2.5cm}{Currency is null}}
& Localization Accuracy & \rating{3} Basic & \rating{4} Precise & \rating{5} Complete \\
& RCA Depth & Shallow (code only) & Medium (business logic) & Deep (configuration) \\
& Runtime Data Completeness & Missing key params & Partially inferred data & Complete actual data \\
& Impact Assessment & Simple exclusion & Not addressed & Identifies upstream factors \\
& Operability & Low & Medium & High \\
\cmidrule{1-5} 
\multirow{4}{*}{\parbox{2.5cm}{The riskcloud policy engine returned anti-fraud flag with REJECT}}
& Raw Evidence Collection & \rating{5} Complete & \rating{4} Mostly Complete & \rating{5} Complete \\
& Causal Chain Analysis & \rating{2} Lacking & \rating{4} Clear & \rating{5} In-depth \\
& Code Localization & \xmark~Failure & \rating{4} Partially successful & \rating{5} Precise \\
& Technical Depth & \rating{2} Shallow & \rating{4} Medium & \rating{5} In-depth \\
\cmidrule{1-5}

\multirow{4}{*}{\parbox{2.5cm}{Error code mapping wrong rules}}
& Config. Query Mechanism & \xmark~Lacking & \xmark~Lacking & \cmark~Detailed query flow \\
& Fallback Logic & \xmark~Not addressed & \xmark~Not addressed & \cmark~Explicit error handling \\
& Config. Storage Location & \xmark~Vague & \xmark~Vague & \cmark~Pinpoints \texttt{DATA\_ITEM\_NAME} \\
& Causal Chain Depth & \dmark~Shallow & \dmark~Medium & \cmark~Deep analysis \\
\cmidrule{1-5}

\multirow{3}{*}{\parbox{2.5cm}{Security controller passed a REJECT decision}}
& Architecture Understanding & \xmark~Assumes external deps & \warn~Single-module analysis  & \cmark~Module-chain arch  \\
& Causal Chain Completeness & \xmark~Breaks at external services & \warn~Lacks normalization step  & \cmark~Complete closed-loop  \\
& Operability & \xmark~Requires external team  & \warn~Partially operable  & \cmark~Internally diagnosable  \\
\cmidrule{1-5}

\multirow{5}{*}{\parbox{2.5cm}{Page is oversize}}
& Default Threshold ID & \cmark~100 & \xmark~200 & \cmark~100 \\
& Core Code Localization & \xmark~Lacking & \cmark~Complete & \cmark~Complete \\
& Error Chain Completeness & \xmark~Incomplete & \cmark~Complete & \cmark~Complete \\
& Log Evidence & \xmark~Lacking & \cmark~Complete & \cmark~Complete \\
& Issue Focus & \xmark~Off-topic & \cmark~Highly Focused & \cmark~Highly Focused \\
\cmidrule{1-5}

\multirow{8}{*}{\parbox{2.5cm}{Invalid member status}}
& Report Format & Strict format, categorized by subject & Structured analysis, causal chain narrative & JSON output, organized by field \\
& Technical Depth & \rating{2} Shallow reference & \rating{4} Complete tech. chain & \rating{3} Medium-deep \\
& Evidence Completeness & \rating{3} Basic evidence complete & \rating{4} Code + logs covered & \rating{3} Rich but redundant \\
& Operability & \rating{2} Needs secondary analysis & \rating{4} Points to solution path & \rating{3} Info overload, needs focus \\
& Key Findings & \parbox{3cm}{\textbullet~Basic code localization \newline \textbullet~Monitoring data normal} & \parbox{3cm}{ \textbullet~3-step data transform \newline \textbullet~Specific code line (40) \newline \textbullet~Complete anomaly propagation} & \parbox{3cm}{\textbullet~Cross-stack dependency \newline \textbullet~Physical location info \newline \textbullet~Component version info} \\
& SRE Practicality & \rating{2} Insufficient info & \rating{4} Best practice standard & \rating{3} Information overload \\
& Core Issue Focus & \rating{2} Focused but shallow & \rating{4} Precisely checks FREEZE state & \rating{4} Effectively filters irrelevant info \\
& Recovery Support & \rating{2} Needs more analysis & \rating{4} Directly supports solution & \rating{3} Partially useful info \\

\cmidrule{1-5}

\multirow{5}{*}{\parbox{2.5cm}{Contract version issue}}
  &Localization Accuracy   & \rating{3} Basic         & \rating{4} Precise       & \rating{5} Complete        \\
&RCA Depth               & Shallow (code only)      & Medium (business logic)  & Deep (configuration)       \\
&Runtime Data Completeness & Missing key params     & Partially inferred data  & Complete actual data       \\
&Impact Assessment       & Simple exclusion         & Not addressed            & Identifies upstream factors\\
&Operability             & Low                      & Medium                   & High                       \\

\bottomrule
\end{tabular}
\end{sidewaystable}

To answer \textbf{RQ3} and quantify the contribution of our core architectural mechanisms, we conducted a qualitative ablation study. The evaluation, detailed in Table~\ref{tab:comprehensive_comparison}, is grounded in a set of representative business-logic-related cases selected from our \texttt{AntRCA} (used in Case Studies 1 and 2) and live production incidents. The performance across each dimension was calibrated by SRE experts to ensure a consistent and accurate assessment.

\paragraph{V1 (Basic ReAct): A Brittle Baseline}
The V1 agent, representing a standard ReAct implementation, served as our baseline. As the table shows, its capabilities are severely limited. In the ``Currency is null'' case, its analysis is ``Shallow (code only),'' and for the ``Riskcloud policy'' case, it completely fails at ``Code Localization.'' This confirms that a simple, unstructured agent loop is insufficient for complex industrial tasks; it lacks the depth to understand business logic and the robustness to navigate multi-step causal chains.

\paragraph{V2 (Phased ReAct): The Value of Structured Reasoning and Experience}
The introduction of a predefined, phased workflow in V2 yields significant improvements. This enhancement is driven by two key mechanisms: first, V2 emulates a developer's code walkthrough experience by generating an agent that performs taint analysis to explore potential data flow issues; second, its reasoning is augmented with domain-specific prompts optimized for business logic. These additions allow V2 to deliver a ``Structured analysis'' and follow a ``Complete tech. chain'' in the ``Invalid member status'' case, a capability far beyond V1's shallow analysis. However, V2's monolithic nature remains a bottleneck. It struggles with cross-domain analysis, as shown in the ``Currency is null'' case where it fails to perform an ``Impact Assessment,'' and in the ``Security controller'' case, it is confined to ``Single-module analysis.'' This highlights the limitations of a single agent trying to be a generalist.

\paragraph{V3 (Multi-Specialist): The Critical Contribution of Collaborative Specialization}
The transition to the V3 Multi-Specialist framework represents the most significant leap, directly addressing the limitations of the V2 agent. The ``divide and conquer'' paradigm proves to be the key mechanism for achieving production-grade analysis:
\begin{itemize}
    \item \textbf{Depth and Breadth:} In the ``Currency is null'' case, V3 achieves ``Deep (configuration)'' analysis and ``Identifies upstream factors.'' This is possible because the Supervisor agent can delegate tasks to specialized agents (e.g., a \texttt{Data-Agent} for runtime data and a \texttt{Config-Agent} for configuration parameters), enabling a holistic, cross-stack view that a single agent cannot manage.

    \item \textbf{Architectural Awareness:} For the ``Security controller'' case, V3 demonstrates a true understanding of the ``Module-chain arch.,'' creating a ``Complete closed-loop'' analysis. This capability stems directly from the multi-agent collaboration, which can trace dependencies across service boundaries.

    \item \textbf{Precision and Focus:} In the ``Error code mapping'' case, V3 is able to execute a ``Detailed query flow'' and ``Pinpoints \texttt{DATA\_ITEM \_NAME},'' showcasing the power of specialized tool use within the framework. Similarly, in the ``Page is oversize'' case, V3 remains ``Highly Focused'' where V1 was ``Off-topic,'' demonstrating the Supervisor's role in maintaining task coherence.
\end{itemize}

\paragraph{Conclusion for RQ3}
Our analysis confirms that while a structured single-agent (V2) provides a solid foundation, multi-specialist collaboration (V3) is the critical architectural leap that delivers the robustness and depth essential for enterprise-grade diagnostics.

\section{Discussion and Limitations}
Despite its promising results, \derisk has several key limitations. First, it operates as an assistive "co-pilot" requiring human oversight for final remediation, rather than a fully autonomous system. Second, its diagnostic accuracy is fundamentally dependent on the quality and maintenance of the underlying knowledge base. Third, there is an inherent trade-off between analytical depth and performance, as the most accurate multi-agent approach incurs higher latency and computational cost. Finally, its generalizability beyond the specific and mature observability ecosystem of Ant Group has not yet been validated in organizations with different technology stacks or data quality.
\section{Conclusion and Future Work}
\label{sec:conclusion}

In this paper, we introduced \derisk, a general-purpose AI-SRE framework designed not only for complex Root Cause Analysis (RCA) but also to empower developers to autonomously create new agents for a wide range of SRE tasks. Our primary contribution is the demonstration that a multi-specialist collaboration paradigm (V3) significantly outperforms monolithic agent architectures in diagnostic depth, accuracy, and robustness.

By integrating a sophisticated Knowledge Engine, adaptive collaboration patterns, and a pluggable Reasoning Engine, \derisk provides a practical and powerful ``co-pilot'' for SRE teams. This practicality and scalability are substantiated by its successful production deployment at Ant Group. Over a three-month period, the framework has been adopted for \textbf{13 new application scenarios}, with developers creating over \textbf{50 new specialized agents}. Today, it serves more than \textbf{3,000 daily users} and executes over \textbf{60,000 runs per day}. This real-world data, combined with our experimental results, validates our design philosophy and marks a significant step towards building truly intelligent, adaptable, and scalable diagnostic systems for enterprise environments.

Our roadmap focuses on evolving \derisk from a "co-pilot" to an autonomous "pilot" (V4) capable of executing safe, closed-loop remediation. This will be driven by a comprehensive Reinforcement Learning paradigm to optimize both individual agent tool-use and system-level collaboration strategies. We also plan to develop a self-healing Knowledge Engine, where agents contribute to their own knowledge base, and an adaptive reasoning controller to dynamically balance diagnostic depth against response latency.

\clearpage
\bibliographystyle{colm2024_conference}
\bibliography{ref}

\appendix
\section{Contributions and Acknowledgments}

\begin{table}[h]
    \centering
    \begin{tabular}{p{0.4\linewidth}p{0.4\linewidth}}
Ang Zhou	&	Peng Di	\\
Bingchang Liu	&	Peng Tang	\\
Changle Zhang	&	Qi Zhang	\\
Cheng Zeng	&	Qiao Su	\\
Faqiang Chen	&	Qing Ouyang	\\
Fei Gao	&	Qingfeng Li	\\
Feng Shi	&	Sheng Gao	\\
Ganglin Wei	&	Shenglong Jing	\\
Guilin Li	&	Silin Hu	\\
Hang Yu	&	Songshan Luo	\\
Hao Jiang	&	Tingting Li	\\
Hao Liu	&	Wenfei Li	\\
Hongchao Le	&	Wenhui Shi	\\
Hongjun Yang	&	Xiao Bai	\\
Jian Mou	&	Yali Wang	\\
Jianfei Zhang	&	Yaodong Li	\\
Jie Bao	&	Yihui He	\\
Jie Yang	&	Yin Hu	\\
Jingwei Qu	&	Yujing Zhang	\\
Junwei Guo	&	Yun Wang	\\
Keting Chen	&	Yunfeng Chen	\\
Linxia Zhong	&	Zheng Li	\\
Mao Ren	&	Zhitao Shen	\\
Min Shen	&	Zilong Hou	\\
Lei Lei	&		\\
    \end{tabular}
\end{table}



\end{document}